\documentstyle[12pt]{article}
\begin{document}
\bibliographystyle{unsrt}

\title{The Signatures Of Glueballs In $J/\psi$ Radiative Decays}
\author{Zhenping Li \\
Physics Department, Peking University \\
Beijing, 100871, P.R. China}

\maketitle

\begin{abstract}
In this talk, I shall discuss the signatures of glueballs 
in the $J/\psi$ radiative decays. Further experimental and
 theoretical investigations are suggested. 
\end{abstract}

\vskip .5cm

The existence of glueballs and hybrids in nature has been one of the 
important predictions of the quantum chromodynamics(QCD).  Considerable 
progress has been made recently in identifying the glueball candidates. 
In the scalar meson sector, the $f_0(1300)$, $f_0(1500)$\cite{cbr} and
$f_0(1780)$\cite{bes96} have been established in the recent experiments 
and this has raised the possibility that these three scalars are the 
mixed states between the ground state glueball and two nearby $q\bar q$ nonet.  
The study has shown\cite{amc} that the observed properties of these states 
are incompatible with them being $q\bar q$ states, and one of them, in particular $f_0(1500)$, might be a ground state glueball.  Moreover,
the discovery\cite{bes} of non-strange decay modes of the 
state $\xi(2230)$ in addition to the strange decay channel observed 
in earlier experiments have also fueled the speculation of it being a 
tensor glueball state. The observed relative strength of each
decay mode of the  $\xi(2230)$ shows a remarkable flavor symmetry, which is
one of the important characteristics of a glueball state.  In this talk, 
I shall concentrate on the theoretical and experimental aspects of 
the $J/\psi$ radiative decays,
which has become increasingly important in identifying the glueball 
candidates.  

The advantage of studying the glueball productions in $J/\psi$
radiative decays is that the properties of the glueballs can be 
investigated not only via their decays but also through their
productions.  According to the perturbative QCD, the production 
of a light meson state $R$ in the $J/\psi$ radiative decay proceeds by the 
sequence $J/\psi \to \gamma + gg\to \gamma + R$. In leading order pQCD, 
its amplitude $A$ is given by
\begin{equation}\label{QQgamR}
A= \frac 12\sum \int \frac {d^4k}
{(2\pi)^4} \frac 1{k_1^2} \frac 1{k_2^2} <(Q\bar Q)_V | \gamma g^ag^b> 
<g^ag^b | R>.
\end{equation}
The summation is over the polarization vectors $\epsilon_{1,2}$ and
color indices $a,b$ of the intermediate gluons, whose momenta are
denoted as $k_{1,2}$.  Thus, there are three major components in evaluating
the $J/\psi$ radiative decays; the inclusive process $J/\psi \to
\gamma + gg$ whose amplitude $<(Q\bar Q)_V | \gamma g^ag^b>$ has been 
given reliably in pQCD, the process $gg\to R$ and the loop integral.  
The process $gg\to R$ for a glueball state has not been investigated before.
We find\cite{cfl} that it is reasonable to assume 
 the amplitude $<g^ag^b | R>$ for both $q\bar q$ and glueball states
having the form 
\begin{equation}\label{extra3}
\psi(R)=\left \{ \begin{array}{cc} \frac 1{\sqrt{3}} P_{\mu\nu}
{G^{1a}_{\mu\rho} G^{2a}_{\nu\rho}} F_0(k_1^2,k_2^2) & \mbox{for $0^{++}$} \\
\epsilon_{\mu\nu}{G^{1a}_{\mu\rho} G^{2a}_{\nu\rho}} F_2(k_1^2,k_2^2) & 
\mbox{for $2^{++}$} \end{array} \right. 
\end{equation}
 where $P_{\rho\sigma} \equiv g_{\rho\sigma}-\frac {P_{\rho}P_{\sigma}}{m^2}$
for a resonance with mass $m$ and momentum $P_{\mu}$, and
  $\epsilon_{\rho\sigma}$ are the tensor for a tensor
resonance, and satisfy the relations
\begin{equation}\label{vector}
\sum_{\epsilon}
\epsilon_{\rho\sigma}\epsilon_{\rho^\prime\sigma^\prime} 
=  \frac 12(P_{\rho\rho^\prime}P_{\sigma\sigma^\prime}
+P_{\rho\sigma^\prime}P_{\sigma\rho^\prime})-\frac
13P_{\rho\sigma}P_{\rho^\prime\sigma^\prime}.
\end{equation}
A direct consequence from Eq. \ref{extra3} is that 
ratio of the two gluon width between the scalar and the tensor states 
is
\begin{equation}
\frac {\Gamma(0^{++})}{\Gamma(2^{++})}=\frac {15}4 
\end{equation}
for both $q\bar q$ and glueball states assuming equal masses and
 form factors.  Qualitatively one would 
expect that the total width for a tensor glueballs should be of order 
$O(25 MeV)$  if the width for the scalar is at $O(100 MeV)$ suggested 
by the states $f_0(1500)$ and $f_0(1780)$.  Of course, these are 
circumstantial arguments for the $\xi(2230)$ being a tensor glueball as
its total width is around $20\sim 30$ MeV, and the experimental 
determination of the spin of $\xi(2230)$ is calling for. 
 
The form factor $F(k_1^2,k_2^2)$ in Eq. \ref{extra3}
is well established for the $q\bar q$ states, while there is little
information on this form factor for glueball states.  Its determination
for glueball states depends how much we understand the structure of 
their wavefunctions.  
Assuming that the $q\bar q$ and gleuball states have the same 
form factor, the branching ratio for the $J/\psi$ radiative decaying
 into a resonance $R$ with the mass $m$ has a general form\cite{cak};
\begin{equation}\label{f5}
B(J/\psi \rightarrow \gamma +R_J) = B(J/\psi \to \gamma +gg)
C_{R_J}\Gamma(R_J \rightarrow gg) \frac{x|H_J|^2}{8\pi(\pi^2-9)}
 \frac{m}{M^2},
\end{equation}
where $M$ is the mass of the state $J/\psi$ and $x=1-\left 
(\frac mM\right)^2$. The coefficient $C_R$ in Eq. \ref{f5} depends on 
the spin parity of the final resonance $R$, and it is
\begin{equation}\label{f6}
C_{R_J} \equiv 1\ (0^{-+}); \frac 23 \ (0^{++}); \frac 52 \ (2^{++}).
\end{equation}
The quantity $B(J/\psi \to \gamma + gg)$ is a branching ratio for the
inclusive process. It is determined by the vertex $J/\psi \to \gamma gg$,
and its numerical value has been well determined,
which gives $B(J/\psi \to \gamma + gg)\approx 0.06 \sim 0.08$. 
The $H_J(x)$ in Eq. 
\ref{f5} is a loop integral, and it has been evaluated in the case of 
$R=q\bar q$\cite{korner}.
The quantity $\Gamma(R_J\to gg)$ represents the width of the resonance
$R$ decaying into the two gluon state $gg$,  which determines the vertex
$R\to gg$.  Generally the decay of a resonance $R$ into the two gluon state
$gg$ is not the same as its total decay width, since the gluon hadronization
is not the major decay mode for a light $q\bar q$ meson.  Thus, one can define  
a branching ratio $b(R_J\to gg)$ so that $\Gamma(R_J\to gg) =b(R_J\to gg)
\Gamma_T$,  which measures the gluonic content of a resonance $R$.
 Cakir and Farrar\cite{cak} argued that 
\begin{equation}\label{f66}
b(R(q\bar q)\to gg)=O(\alpha_s^2) \simeq 0.1\ \sim \ 0.2
\end{equation}
for a normal $q\bar q$ meson, while 
\begin{equation}\label{f7}
b(R(G)\to gg)\simeq 0.5 \sim 1
\end{equation}
for a glueball state.

It is the branching ratio $b(R\to gg)$ for a resonance $R$ that can be 
extracted from the data for $B(J/\psi \to \gamma +R)$ and $\Gamma_T$ with 
the theoretical input of the loop integral $x|H(x)|^2$ in Eq. \ref{f5}.
The numerical results from Ref. \cite{korner} show that the loop integral
$x|H(x)|^2 \approx 35 \sim 40$ for the scalar and tensor states with masses
around 1.5 GeV in $J/\psi$ radiative decays.  Thus, one can rewrite Eq. 
\ref{f5} as\cite{cfl}
\begin{eqnarray} \label{f11}
10^3  B(\psi \rightarrow \gamma 2^{++}) = \left (\frac{m}{1.5\; GeV}\right )
\left (\frac{\Gamma_{R\rightarrow gg}}{26\; MeV} \right ) 
\frac{x|H_T|^2}{34}\nonumber \\
10^3  B(\psi \rightarrow \gamma + 0^{++}) = \left (\frac{m}{1.5\; GeV}\right )
\left (\frac{\Gamma_{R\rightarrow gg}}{96\; MeV}\right )\frac{x|H_S|^2}{35},
\end{eqnarray}
A straightforward evaluation shows that
 the branching ratios $b(R\to gg)$ extracted from the $J/\psi$ 
radiative decay data for the established $q\bar q$ mesons, such 
as $f_2(1270)$ and $f_2(1525)$, clearly satisfy Eq. \ref{f66}, while 
the existing data for $f_0(1500)$ and $f_J(1710)$ may well be the examples 
of Eq. \ref{f7}.  The particle data group\cite{pdg96} gives 
\begin{equation}\label{42}
B(J/\psi \to \gamma + f_0(1500))=(0.82\pm 0.15)\times 10^{-3},
\end{equation}
which is in good agreement with the recent BES results\cite{bes97} in 
$J/\psi\to \gamma f_0(1500) \to \gamma \pi^0\pi^0$, which translates into
\begin{equation}\label{6}
b(f_0(1500)\to gg)=0.64\pm 0.11, 
\end{equation}
assuming that $\Gamma_{T}\approx 120\pm 20 \mbox{MeV}$.  This suggests that
the resonance $f_0(1500)$ should have a large glueball component
in its wavefunction.
The recent results from the BES group suggested that the $f_J(1710)$ be 
separated into $f_2(1690)$ and $f_0(1780)$ states, and the scalar 
$f_0(1780)$ is consistent with the analysis in Ref \cite{bugg}
in which a scalar with mass 1.75 GeV and $160$ MeV width is reported.  
Its decay into the $4\pi$ channel is 
very large and dominated by the $\sigma\sigma$ contributions.  
The branching ratio $B(J/\psi \to\gamma + f_0(1750))$ is found to be
\begin{equation} 
B(J/\psi\to \gamma f_0(1750))B(f_0(1750)\to 4\pi)\approx 1.0 \times 10^{-3}
\end{equation}
with the total width $160$ MeV. This corresponds to 
\begin{equation}\label{fj}
b(f_0(1710)\to gg)\ge 0.5.
\end{equation}
Thus, Eqs. \ref{6} and \ref{fj} suggest that the ground state 
glueball should be mixed with the nearby $SU(3)$ $q\bar q$ nonet.  The 
experimental consequences of the configuration mixings between the $q\bar q$
nonet and glueballs in the $J/\psi$ radiative decays and the $\gamma \gamma$
collisions have been discussed extensively in Ref. \cite{cfl}.

Now, we examine the $\xi(2230)$ and its implications. 
The data  from BES collaboration\cite{bes} for $J/\psi \to 
\gamma +\xi(2230)$ in the $K\bar K$ and $p\bar p$ channels 
are\cite{bes}
\begin{eqnarray}\label{BES}
B(J/\psi \to \gamma +\xi(2230))B(\xi(2230)\to K^+K^-)=(3.3\pm 2.5)\times 
10^{-5} \nonumber \\
B(J/\psi \to \gamma +\xi(2230))B(\xi(2230)\to K^0_sK^0_s)=(2.7\pm 2.0)\times 
10^{-5} \nonumber \\
B(J/\psi \to \gamma +\xi(2230))B(\xi(2230)\to p\bar p)=(1.5\pm 1.0)\times 
10^{-5} ,
\end{eqnarray}
while the recent results from JETSET\cite{ps185} has set a very strict 
upper limit on $B(\xi\to p\bar p)B(\xi \to K\bar K)$
\begin{equation}\label{ps85}
B(\xi\to p\bar p)B(\xi \to K\bar K)<7.5\times 10^{-5}.
\end{equation}
This gives a lower limit  for the branching ratio $B(J/\psi\to \gamma 
+\xi(2230))$
\begin{equation} 
b(J/\psi\to \gamma \xi(2230))\ge 2.0 \times 10^{-3}.
\end{equation}
Because $\xi(2230)$ has a narrow width, $\Gamma_T=20$ MeV,  the resulting
branching ratio $b(R\to gg)$ from Eq. \ref{f11} would be
\begin{eqnarray}\label{0++}
b(\xi(2230)|J=0^{++} \to gg) \ge 10\pm 5 \nonumber \\
b(\xi(2230)|J=2^{++} \to gg) \ge 2\pm 1 .
\end{eqnarray}
This suggests that 
a scalar $\xi(2230)$ may have already been excluded by the data, 
while a tensor $\xi(2230)$ is still possible considering the uncertainties
in our approach. To clarify these questions requires the measurements with better statistics in both $p\bar p \to K\bar K$ and $j/\psi \to \gamma 
\xi(2230) \to \gamma p\bar p$.   Eq. \ref{ps85} also suggests that the two body
final states are not the major decay modes for the $\xi(2230)$, and the recent
analysis in $J/\psi\to \gamma 4\pi$ shows\cite{zhu} a $f_2(2220)$ state,
 whose decay is dominantly via $f_2(1270)\sigma$.  Further analysis 
with better statistics are needed to confirm that $f_2(2220)$ is indeed 
$\xi(2230)$. The analysis $J/\psi\to\gamma K\bar K\pi\pi$ channel would 
be important, as the flavor symmetry would also suggest that $\xi(2230)\to 
f_2(1525)\sigma$ would be another important channel.

Theoretically, the remaining question is the theoretical uncertainties of the
loop integral $x|H(x)|^2$ in Eq. \ref{f5}.  The form factor $F(k_1^2,k_2^2)$
for glueballs is still unknown, and to obtain it requires better knowledge 
of glueball wavefunctions. Another source of such uncertainty is the relativistic effects in $R(q\bar q)\to gg$, which was shown\cite{zpli}
 to be very important for light quark mesons.  Thus, a lot of more
 theoretical and experimental works remains to be done to understand the 
nature of glueballs and their differences with the normal $q\bar q$ states,
which in turn will help us to identify the glueball states with more confidence.

\subsection*{Acknowledgment} 
The extensive discussions with F. Close, G. Farrar, C.
Meyer, Zhu Yucan and Shen Xiaoyan are gratefully acknowledged.  
 This work was supported in part by Peking University.


\begin{thebibliography}{99}
\bibitem{cbr} V.V. Anisovich {\it et al.}, Phys. Lett. {\bf 323B},
233(1994); C. Amsler {\it et al.}, {\it ibid.} {\bf 342B}, 433(1994); {\bf
291B}, 347(1992); {\bf 340B}, 259(1994).
\bibitem{bes96} BES Collaboration, J.Z. Bai {\it et al.}, Phys. Rev. Lett.
{\bf 77}, 3959(1996).
\bibitem{amc}C. Amsler and F.E.Close, Phys Lett {\bf B353} (1995) 385;
Phys Rev {\bf D53}, 295(1996).
\bibitem{bes} J. Bai {\it et al.}, BES collaboration, Phys. Rev. Lett.
{\bf 76}, 3502(1996).
\bibitem{cfl} F. E. Close, G. Farrar and Zhenping Li, Phys. Rev. {\bf D55}, 
5749(1997).
\bibitem{korner}J. G. K\"orner, J.H.K\"uhn, M. Krammer, and H. Schneider,
Nucl. Phys. {\bf B229}, 115(1983), J. G. K\"oner, J. H. K\"uhn, and H.
Schneider,  Phys. Lett. {\bf 120B},  444(1983).
\bibitem{cak} M.B. Cakir and G. Farrar, Phys. Rev. {\bf  D50}, 3268(1994).
\bibitem{pdg96} Particle Data Group, Phys. Rev. {\bf D54}, 1(1996).
\bibitem{ps185} A. Buzzo, {\it et al.}  JETSET Collaboration, to be published.
\bibitem{bugg} D. V. Bugg, {\it et al.}, Phys. Lett. {\bf 353B}, 378(1995).
\bibitem{bes97} BES Collaboration, to be published.
\bibitem{zpli} Zhenping Li, F. E. Close and T. Barnes, Phys. Rev. {\bf
D43}, 2621(1991).
\bibitem{zhu} Zhu Yucan, private communication.
\end{thebibliography}
\end{document}